\documentclass[aps,final,prd,eqsecnum,showpacs,letterpaper]{revtex4}
\usepackage{graphicx}
\usepackage[tight,TABTOPCAP]{subfigure}
\usepackage{amsmath}
\usepackage{color}

\makeatletter
\newbox\slashbox \setbox\slashbox=\hbox{$/$}
\newbox\Slashbox \setbox\Slashbox=\hbox{\large$/$}
\def\pFMslash#1{\setbox\@tempboxa=\hbox{$#1$}
  \@tempdima=0.5\wd\slashbox \advance\@tempdima 0.5\wd\@tempboxa
  \copy\slashbox \kern-\@tempdima \box\@tempboxa}
\def\pFMSlash#1{\setbox\@tempboxa=\hbox{$#1$}
  \@tempdima=0.5\wd\Slashbox \advance\@tempdima 0.5\wd\@tempboxa
  \copy\Slashbox \kern-\@tempdima \box\@tempboxa}

\def\miss#1{\ifmmode{/\mkern-11mu #1}\else{${/\mkern-11mu #1}$}\fi}
\makeatother

\def\dal{\,\lower0.9pt\vbox{\hrule \hbox{\vrule height 0.35 cm \hskip 0.3 cm
\vrule height 0.35 cm}\hrule}\,}

\begin{document}

\title{Study of the lepton flavor-violating $Z^{\prime}\to\tau\mu$ decay}

\author{J. I. Aranda$^a$, J. Monta\~no$^b$, F. Ram\'irez-Zavaleta$^a$, J. J. Toscano$^c$, and E. S. Tututi$^a$\footnote{Email: tututi@umich.mx}}

\affiliation{$^a$Facultad de Ciencias F\'{\i}sico Matem\'aticas, Universidad Michoacana de San Nicol\'as de Hidalgo,
Avenida Francisco J. M\'{u}jica S/N, 58060,  Morelia, Michoac\'an, M\'exico\\
$^b$ Departamento de F\'isica, CINVESTAV, Apartado Postal 14-740, 07000, M\'exico, D. F., M\'exico\\
$^c$Facultad de Ciencias F\'{\i}sico Matem\'aticas, Benem\'erita Universidad Aut\'onoma de Puebla, Apartado Postal 1152, Puebla, Puebla, M\'exico}

\begin{abstract}
The lepton flavor violating $Z^{\prime}\to\tau\mu$ decay is studied in the context of several extended models that predict the existence of the new gauge boson named $Z^\prime$. A calculation of the strength of the lepton flavor violating $Z^\prime\mu\tau$ coupling is presented by using the most general renormalizable Lagrangian that includes lepton flavor violation. We used the experimental value of the muon magnetic dipole moment to bound this coupling, from which the $\mathrm{Re}(\Omega_{L\mu\tau}\Omega^\ast_{R\mu\tau})$ parameter is constrained and it is found that $\mathrm{Re}(\Omega_{L\mu\tau}\Omega^\ast_{R\mu\tau})\sim 10^{-2}$ for a $Z^\prime$ boson mass of $2$ TeV. Alongside, we employed the experimental restrictions over the $\tau\to\mu\gamma$ and $\tau\to\mu\mu^+\mu^-$ processes in the context of several models that predict the existence of the $Z^\prime$ gauge boson to bound the mentioned coupling. The most restrictive bounds come from the calculation of the three-body decay. For this case, it was found that the most restrictive result is provided by a vector-like coupling, denoted as $|\Omega_{\mu\tau}|^2$, for the $Z_\chi$ case, finding around $10^{-2}$ for a $Z^\prime$ boson mass of $2$ TeV. We used this information to estimate the branching ratio for the $Z^\prime\to\tau\mu$ decay. According to the analyzed models the least optimistic result is provided by the Sequential $Z$ model, which is of the order of $10^{-2}$ for a $Z^\prime$ boson mass around $2$ TeV.

\end{abstract}

\pacs{ 12.60.Cn, 11.30.Hv, 14.70.Pw
\\ {\em Keywords:} Flavor violation,  Lepton physics
}
\vskip2.5pc
\maketitle

\section{Introduction}

Besides the search for the Higgs boson or  the quark-gluon plasma, experiments in the LHC are also dedicated to looking for clues on the origin of the masses and extra dimensions~\cite{Franceschini,cms2}. Experiments are also focused to finding new particles beyond the standard model (SM) such as magnetic monopoles, strangelets, SUSY particles \cite{ellis1,atlas1,cms1}, etc. In fact, the recent results by the CMS and ATLAS collaborations~\cite{cms1a,cms3,cms4,atlas2}, do not exclude the existence of primed massive gauge bosons ($Z^{\prime}$ or $W^{\prime}$) for a mass above the energy of 2 TeV approximately. In particular, the existence of  $Z^{\prime}$ gauge boson is excluded for a mass below of 1.94 TeV with a 95\% confidence level by the CMS Collaboration~\cite{cms1a}. Moreover, the ATLAS Collaboration establishes a lower limit of 2.21 TeV on the $Z^{\prime}$ gauge boson mass~\cite{atlas2}.  The existence of the $Z^{\prime}$ boson is predicted in  several  extensions of the SM~\cite{pleitez,langacker1,leike,perez-soriano,langacker-rmp}.  The simplest  one is that which it simply adds to the SM  an extra  gauge symmetry group  $U^{\prime}(1)$~\cite{langacker-rmp,leike,robinett,langacker2,arhrib}.

For the SM, flavor changing neutral currents (FCNC) transitions in the quark sector are allowed, but they are very suppressed by the GIM mechanism and because they  emerge at the one loop level. However,  these transitions  could be enhanced greatly by new physics effects produced by the $Z^{\prime}q_iq_j$ couplings where $q_i$ and $q_j$ are up- or down quarks \cite{arhrib,arandaetal}. On the other hand, in the lepton sector, the SM Lagrangian has an exact lepton  flavor  symmetry. Although experimental data show that this symmetry is not satisfied, since it  has already been evidenced in different situations the neutrino oscillations, in which only the total leptonic number is conserved~\cite{neutrinos}. Calculation of transitions between charged leptons is reasonably simpler than  calculations of transitions  in the sector of quarks, since the former leads us without the complications of the CKM elements nor the complications  of the QCD elements of matrix.

Transitions between charged leptons is an important issue since,  if transitions between charged leptons occur, it will be a clear signal of lepton flavor violation (LFV). In the minimally extended  standard model, at  which it is added right-handed neutrinos, the branching ratio $\mathrm{Br}(\tau\to\mu\gamma)$ goes as mass of the neutrino over the $W$ gauge boson mass to the forth power, it results   less than $10^{-40}$~\cite{cheng-li} which is very far from the present experimental capabilities.  This value is such that, when compared with the present experimental data bound  given in the particle data group~\cite{pdg}, namely $\mathrm{Br}(\tau\to\mu\gamma)< 10^{-8}$, is not so restrictive as the obtained from the minimally extended SM. In order to make the theoretical value of  branching ratio  consistent with the experimental result it is necessary to go into a theory beyond the SM which account the LFV. There exist several theories beyond the SM that predict LFV~\cite{langacker-rmp,pleitez,lfv}. However, the simplest  one is  provided by making minimal extensions of the SM~\cite{langacker-rmp}.

In this work we  calculate a bound for the coupling ${Z^{\prime}\tau\mu}$ by using the most general renormalizable lagrangian that includes lepton flavor violation mediated by a new neutral massive gauge boson (namely $Z^\prime$ gauge boson). We  bound  the  mentioned coupling in the spirit of a model independent approach (since no assumptions or  extra-parameters are required) by means of  the current experimental value of the magnetic dipole moment of the $\mu$-lepton and the experimental restrictions for the lepton flavor-violating $\tau\to\mu\gamma$ and $\tau\to\mu\mu^+\mu^-$ decays. For the $\tau\to\mu\gamma$ and $\tau\to\mu\mu^+\mu^-$ decays transitions, we calculated it by resorting to different grand unification theories (GUT)~\cite{langacker-rmp, arhrib}. Then we determine bounds for the  branching ratio of the process $Z^{\prime}\to \tau\mu$ in the same context of GUT's used.

\section{Bounding the $Z^\prime\tau\mu$ coupling}

In order to calculate the branching ratio of $Z^\prime \to  \tau\mu$ decay process, we employ the most general renormalizable Lagrangian that includes lepton flavor violation mediated by a new neutral massive gauge boson, coming from any extended or grand unification model~\cite{durkin, langacker3, Salam-Mohapatra}, which can be cast in the following way
\begin{equation}\label{lnc}
\mathcal{L}_{NC}=\sum_{i,j}\left[\, \overline{f}_i\,
\gamma^{\alpha} (\Omega_{{L}f_if_j} \, P_L+\Omega_{{R}f_if_j} \,
P_R)\, f_j+\overline{f}_j\, \gamma^{\alpha} ({\Omega^\ast_{{L}f_jf_i}}\,
P_L+{\Omega^\ast_{{R}f_jf_i}} \, P_R) \, f_i \, \right]Z^{\prime}_{\alpha},
\end{equation}
where $f_{i}$ is any fermion of the SM, $P_{L,R}=\frac{1}{2}(1\pm \gamma_{5})$ are the chiral projectors, and $Z^\prime_\alpha$ is a new neutral massive gauge boson predicted by several extensions of the SM~\cite{durkin, langacker3, Salam-Mohapatra, Pleitez}. The $\Omega_{Ll_il_j}$, $\Omega_{Rl_il_j}$ parameters represent the strength of the $Z^\prime l_il_j$ coupling, where $l_i$ is any charged lepton of the SM. In the rest of the paper we will assume that $\Omega_{Ll_il_j}=\Omega_{Ll_jl_i}$ and $\Omega_{Rl_il_j}=\Omega_{Rl_jl_i}$. The Lagrangian in Eq.~(\ref{lnc}) includes both flavor-conserving and flavor-violating couplings mediated by a $Z^\prime$ gauge boson. This work is oriented to study the impact of lepton flavor-violating couplings mediated by a $Z^\prime$ boson in the $Z^\prime\to \tau\mu$ decay. For this purpose we need to bound the lepton flavor-violating coupling $Z^\prime\tau\mu$. This task will be realized by using the experimental result of the muon anomalous magnetic dipole moment and the experimental restrictions for the lepton flavor-violating $\tau\to\mu\gamma$ and $\tau\to\mu\mu^+\mu^-$ decays. Moreover, in order to calculate the branching ratio of the $Z^\prime\to \tau\mu$ process it is necessary to know the $Z^\prime$ total width decay, which mainly depends on the $Z^\prime\to f_i\bar{f}_i$ decays. These flavor-conserving couplings are model dependent.

Here, we only consider the following $Z^\prime$ bosons: the $Z_S$ of the sequential $Z$ model, the $Z_{LR}$ of the left-right symmetric model, the $Z_\chi$ arising from the breaking of $SO(10)\to SU(5)\times U(1)$, the $Z_\psi$ resulting in $E_6\to SO(10)\times U(1)$, and the $Z_\eta$ appearing in many superstring-inspired models~\cite{langacker2}. Concerning to the flavor-conserving couplings, $Q^{f_i}_{L,R}$~\cite{robinett,langacker2,arhrib}, whose values are shown in Table~\ref{table1} for different extended models are related to the $\Omega$ couplings as $\Omega_{Lf_if_i}=-g_2 \,Q^{f_i}_L$ and $\Omega_{Rf_if_i}=-g_2 \,Q^{f_i}_R$, where $g_2$ is the gauge coupling of the $Z^\prime$ boson. For the various extended models we are interested, the gauge couplings of $Z^\prime$'s are
\begin{equation}
g_2=\sqrt{\frac{5}{3}}\sin\theta_W g_1\lambda_g,
\end{equation}
where $g_1=g/\cos\theta_W$. $\lambda_g$ depends of the symmetry breaking pattern being of $\mathcal{O}(1)$~\cite{robinett2} and $g$ is the weak coupling constant. In the sequential $Z$ model, the gauge coupling $g_2=g_1$.
\begin{table}[htb!]
\caption{\label{table1}
Chiral-diagonal couplings of the extended models.}
\medskip
\begin{ruledtabular}
\begin{tabular}{cccccc}
      & $Z_S$ & $Z_{LR}$ & $Z_\chi$ & $Z_\psi$ & $Z_\eta$ \\
\hline
$Q_L^u$ & $0.3456$ & $-0.08493$ &  $\frac{-1}{2\sqrt{10}}$ &
      $\frac{1}{\sqrt{24}}$ & $\frac{-2}{2\sqrt{15}}$ \\
$Q_R^u$ & $-0.1544$  & $0.5038$ &  $\frac{1}{2\sqrt{10}}$ &
      $\frac{-1}{\sqrt{24}}$ & $\frac{2}{2\sqrt{15}}$ \\
$Q_L^d$ & $-0.4228$ & $-0.08493$ &  $\frac{-1}{2\sqrt{10}}$ &
      $\frac{1}{\sqrt{24}}$ & $\frac{-2}{2\sqrt{15}}$ \\
$Q_R^d$ & $0.0772$ & $-0.6736$ &  $\frac{1}{2\sqrt{10}}$ &
      $\frac{-1}{\sqrt{24}}$ & $\frac{2}{2\sqrt{15}}$ \\
$Q_L^e$ & $-0.2684$ & $0.2548$ &  $\frac{3}{2\sqrt{10}}$ &
      $\frac{1}{\sqrt{24}}$ &  $\frac{1}{2\sqrt{15}}$ \\
$Q_R^e$ & $0.2316$ & $-0.3339$ &  $\frac{-3}{2\sqrt{10}}$ &
      $\frac{-1}{\sqrt{24}}$ & $\frac{-1}{2\sqrt{15}}$ \\
$Q_L^\nu$ & $0.5$ & $0.2548$ &  $\frac{3}{2\sqrt{10}}$ &
      $\frac{1}{\sqrt{24}}$ &  $\frac{1}{2\sqrt{15}}$ \\
\end{tabular}
\end{ruledtabular}
\end{table}
\subsection{Muon anomalous magnetic dipole moment}
The contribution of the $Z^\prime \tau\mu$ vertex to the muon anomalous magnetic dipole moment is given through the Feynman diagram shown in Fig.~\ref{Figuras1}.

The flavor changing amplitude for the on-shell $l_i l_i \gamma$ vertex can be written as follows:

\begin{figure}[ht]
\centering
\includegraphics[scale=0.8]{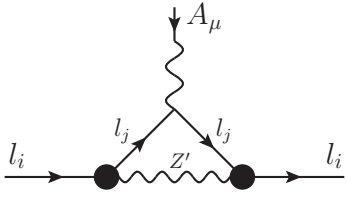}
\caption{Diagram contributing to the magnetic and electric dipole moments of the $l_i$ fermion.}
\label{Figuras1}
\end{figure}

\begin{equation}\label{amp1}
\mathcal{M}=\overline{u}(p_2)\,\Gamma_{\mu}\, u(p_1)\, \epsilon^{\ast \mu}(q,\lambda),
\end{equation}

where the vertex function is given by
\begin{equation}\label{amp2}
\Gamma_{\mu} =e\, \int \frac{d^Dk}{(2\pi)^D}\, \frac{T_\mu}{\Delta},
\end{equation}

with
\begin{align}\label{amp3}
T_{\mu}=& \big (\, \gamma^\alpha\, (\pFMSlash{k}+\pFMSlash{p_2}) \gamma_\mu (\pFMSlash{k}+\pFMSlash{p_1}) \gamma_\alpha   +m_j^2  \gamma^\alpha \gamma_\mu \gamma_\alpha\, \big )\, \big ( G_{Vij}+G_{Aij}\, \gamma_5 \big )+ \,4\, m_j \, \big( 2\, k_\mu +p_{1\mu}+p_{2\mu}\big)\, \big( C_{Vij}+ C_{Aij}\,\gamma_5 \big) ,\\
\Delta=&\big( k^2-m_{Z^\prime}^2 \big)\, \big( (k+p_1)^2-m_j^2 \big)\, \big( (k+p_2)^2-m_j^2 \big),
\end{align}
where
\begin{align}\label{amp4}
G_{Vij}&=\frac{|\Omega_{Lij}|^2+|\Omega_{Rij}|^2}{2},\\
G_{Aij}&=\frac{|\Omega_{Lij}|^2-|\Omega_{Rij}|^2}{2},\\
C_{Vij}&=\frac{\Omega_{Lij}\,\Omega_{Rij}^{\ast}+\Omega_{Rij}\,\Omega_{Lij}^{\ast}}{2},\\
C_{Aij}&=\frac{\Omega_{Rij}\,\Omega_{Lij}^{\ast}-\Omega_{Lij}\,\Omega_{Rij}^{\ast}}{2}.
\end{align}

After using the well known Gordon identities, it is easy to see that there are contributions to the monopole $\big[G(q^2)\,\gamma_\mu\big]$, the magnetic dipole moment $\big[i\,(a_i/2\,m_i)\,\sigma_{\mu \nu}\, q^\nu \big]$, and the electric dipole moment $(-d_i\, \gamma_5\,\sigma_{\mu \nu}\, q^\nu )$. The contribution to the monopole is divergent, but we only are interested in the magnetic and electric dipole moments, for which the contribution is free of ultraviolet divergences. After some algebra, the form factors associated with the electromagnetic dipoles of $l_i$ can be written as follows:

\begin{align}\label{amp5}
a_{i} &=\frac{x_i^2}{4\,\pi^2}\, \big[ (|\Omega_{Lij}|^2+|\Omega_{Rij}|^2) \,f(x_i,x_j)-2\,\frac{x_j}{x_i}\, \mathrm{Re}(\Omega_{Lij}\,\Omega_{Rij}^{\ast})\,g(x_i,x_j)\big],\\
d_i&=\frac{e\,x_j}{4\,\pi^2\,m_{Z^\prime}}\,\mathrm{Im}(\Omega_{Lij}\,\Omega_{Rij}^{\ast})\,\,g(x_i,x_j),
\end{align}

where
\begin{align}\label{amp6}
f(x_i,x_j) &=\int_0^1dx\,\int_0^{1-x}dy\,\frac{(1-x-y)^2}{(1+x_i^2-x_j^2)(x+y)-x_i^2\,(x+y)^2-1},\\
g(x_i,x_j)&=\int_0^1dx\,\int_0^{1-x}dy\,\frac{1-2x-2y}{(1+x_i^2-x_j^2)(x+y)-x_i^2\,(x+y)^2-1}.
\end{align}
In the above expressions the dimensionless variables $x_i=m_i/m_{Z^\prime}$ and $x_j=m_j/m_{Z^\prime}$ have been introduced.

If we take $m_i=m_\mu$, and $m_j=m_e,m_\mu,m_\tau$ and after numerical evaluation it can be appreciated that $f(x_\mu, x_j)$ is suppressed with respect to $g(x_\mu, x_j)$ for about three orders of magnitude. Consequently, the $x_\mu^2f(x_\mu,x_j)$ term in Eq.~(\ref{amp5}) can be neglected, as it is irrelevant compared with $x_\mu x_jg(x_\mu,x_j)$. In particular, the result $x_\mu x_\tau g(x_\mu,x_\tau)\sim 10\, x_\mu x_e g(x_\mu,x_e)$ holds in the interval $1 \, \mathrm{TeV} \leq m_{Z^\prime}\leq 3 \, \mathrm{TeV}$, thus we can neglect the electron contribution. Therefore, we can write only the dominant contribution given by
\begin{align}\label{amp61}
a_{\mu} &=-\frac{x_\mu x_\tau}{2\,\pi^2}\,\mathrm{Re}(\Omega_{L\mu \tau}\,\Omega_{R\mu \tau}^{\ast})\,g(x_\mu,x_\tau)\\
\label{amp62}
d_\mu&=\frac{e\,x_\tau}{4\,\pi^2\,m_{Z^\prime}}\,\mathrm{Im}(\Omega_{L\mu \tau}\,\Omega_{R\mu \tau}^{\ast})\,\,g(x_\mu,x_\tau).
\end{align}
The anomalous magnetic moment of the muon is one of the physical observable best measured. We will use the experimental uncertainty on this quantity to bound $\mathrm{Re}(\Omega_{L\mu \tau}\,\Omega_{R\mu \tau}^{\ast})$. We will assume that the one-loop contribution of the $Z^\prime \mu \tau$ vertex to $a_\mu$ is less than the experimental uncertainty, which is~\cite{pdg}
\begin{equation}\label{amp7}
|\Delta a_\mu^{Exp}|< 6\times 10^{-10}.
\end{equation}
As $\mathrm{Im}(\Omega_{L\mu \tau}\,\Omega_{R\mu \tau}^{\ast})$ is concerned, we can use the existing experimental limit on the muon electric dipole moment, which is~\cite{pdg}
\begin{equation}\label{amp8}
|d_\mu^{Exp}|< 0.1\times 10^{-19} \, \mathrm{e\, cm}.
\end{equation}
\noindent From expressions given by Eqs.~(\ref{amp61}) and~(\ref{amp62}), one can write
\begin{align}\label{amp91}
|\mathrm{Re}(\Omega_{L\mu \tau}\,\Omega_{R\mu \tau}^{\ast})| &<2\pi^2 \left|\frac{\Delta a_\mu^{Exp}}{x_\mu x_\tau \,g(x_\mu,x_\tau)}\right|,\\
\label{amp92} |\mathrm{Im}(\Omega_{L\mu \tau}\,\Omega_{R\mu \tau}^{\ast})|&<4\,\pi^2\left|\frac{d_\mu^{Exp}}{x_\tau \,g(x_\mu,x_\tau)}\right|.
\end{align}

\subsection{The two-body $\tau\to\mu\gamma$ decay}
\begin{figure}[htb!]
\centering
\includegraphics[width=4.5in]{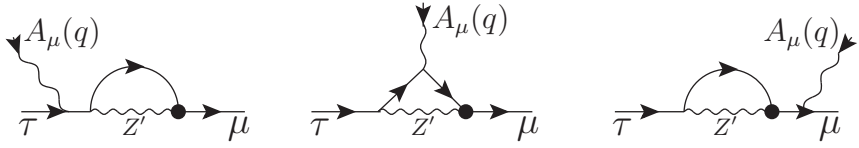}
\caption{\label{tmg} Feynman diagrams contributing to the lepton flavor-violating $\tau\to \mu\gamma$ decay.}
\end{figure}

The contribution of the flavor-violating $Z^\prime \tau\mu$ vertex to the $\tau\to \mu\gamma$ decay is given through the diagrams shown in Fig.~\ref{tmg}. Notice that this  transition is of dipolar type and is  model-dependent since the  coupling  $Z^{\prime}l_il_j$ depends on both, the $g_2$ coupling constant and the chiral couplings $Q^e_R$, $Q^e_L$~\cite{arhrib,durkin,arandaetal,langacker2}. The corresponding amplitude can be reduced to
\begin{align}
\mathcal{M}^\alpha(\tau\to \mu\gamma)=\frac{ieg_2}{64\pi^2m_\tau}\,\bar{u}(p_\mu)\,\sigma^{\alpha\beta}\,q_\beta
&\Bigg[F_1(Q^e_L-Q^e_R)(\Omega_{L\mu\tau}-\Omega_{R\mu\tau}) + F_2(Q^e_L\Omega_{L\mu\tau}+Q^e_R\Omega_{R\mu\tau}) + (F_1(Q^e_L-Q^e_R)\nonumber\\
&\times(\Omega_{L\mu\tau}+\Omega_{R\mu\tau}) + F_2(Q^e_L\Omega_{L\mu\tau}-Q^e_R\Omega_{R\mu\tau}))\gamma_5\Bigg]u(p_\tau),
\end{align}
with
\begin{align}
F_1&=x^2_\tau+6\Bigg(\frac{1-3x^2_\tau}{x_\tau^2(1-x_\tau^2)}\ln x_\tau-\frac{\sqrt{1-4x^2_\tau}}{x^2_\tau}\,\mathrm{arcsech}(2x_\tau)+1\Bigg),\nonumber\\
F_2&=2 + \int_0^1 dx\int_0^x dy\frac{4\,x^2_\tau}{(1-x^2_\tau)(x-y)+(1+xy-y^2)x^2_\tau}-\frac{4}{x_\tau^2}\Bigg(\frac{1-3x^2_\tau}{x_\tau^2(1-x_\tau^2)}\ln x_\tau-\frac{\sqrt{1-4x^2_\tau}}{x^2_\tau}\,\mathrm{arcsech}(2x_\tau)+1\Bigg),\nonumber
\end{align}
where $x_\tau=m_\tau/m_{Z^\prime}$, with $m_\tau$ being the mass of the tau lepton. Furthermore, it can be observed that there is only present a magnetic dipole contribution, which is finite. Also, the muon mass has been neglected. After squaring the amplitude, one obtains the associated branching ratio
\begin{equation}\label{br-2}
\mathrm{Br}(\tau\to \mu\gamma)=\frac{\alpha g_2^2}{4096\pi^4}\Big[|F_1(Q^e_L-Q^e_R)+F_2Q^e_L|^2|\Omega_{L\mu\tau}|^2 + |F_1(Q^e_R-Q^e_L)+F_2Q^e_R|^2|\Omega_{R\mu\tau}|^2\Big]\,\frac{m_\tau}{\Gamma_\tau},
\end{equation}
where $\Gamma_\tau$ is the total decay width of the tau lepton. This branching ratio must be less than the experimental constraint $\mathrm{Br_{Exp}}(\tau\to\mu\gamma)<4.4\times10^{-8}$~\cite{pdg,babar}, so this restriction allow us to bound the $|\Omega_{L\mu\tau}|^2$ and $|\Omega_{R\mu\tau}|^2$ parameters.

\subsection{The three-body $\tau\to\mu\mu^+\mu^-$ decay}
\begin{figure}[htb!]
\centering
\includegraphics[width=2.0in]{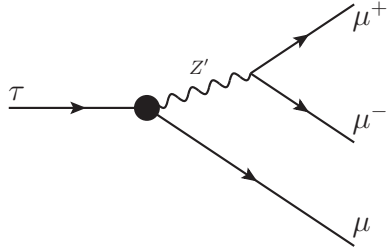}
\caption{\label{tmmm} Feynman diagram corresponding to the $\tau\to\mu\mu^+\mu^-$ decay.}
\end{figure}

We also can compute the contribution of the flavor-violating $Z^\prime \tau\mu$ vertex to the $\tau\to \mu\mu^+\mu^-$ decay (see Fig.~\ref{tmmm}). Since the three-body decay of the tau lepton comes from a tree-level Feynman diagram we only present their corresponding branching ratio

\begin{equation}\label{br3}
Br(\tau\to\mu\mu^+\mu^-)=\frac{g_2^2}{384 \pi^3}(h_1(m_{Z^\prime})(|Q^{e}_{L}\Omega_{L\mu\tau}|^2+|Q^{e}_{R}\Omega_{R\mu\tau}|^2)+h_2(m_{Z^\prime})(|Q^{e}_{L}\Omega_{R\mu\tau}|^2
+|Q^{e}_{R}\Omega_{L\mu\tau}|^2))\,\frac{m_\tau}{\Gamma_\tau},
\end{equation}
where
\begin{align}
h_1(m_{Z^\prime})&=\int_0^1dx\frac{2x-1}{(x-1+m^2_{Z^\prime}/m_\tau^2)}(2(7-4x)x-5),\nonumber\\
h_2(m_{Z^\prime})&=\int_0^1dx\frac{2x-1}{(x-1+m^2_{Z^\prime}/m_\tau^2)}(1-2(x-1)x).\nonumber
\end{align}
The branching ratio computed in Eq.~(\ref{br3}) must be less than the corresponding experimental restriction to the process $\tau\to \mu\mu^+\mu^-$, $\mathrm{Br_{Exp}}(\tau\to \mu\mu^+\mu^-)<2.1\times10^{-8}$~\cite{pdg,belle}, which allow us to get a bound on the flavor-violating $|\Omega_{L\mu\tau}|^2$ and $|\Omega_{R\mu\tau}|^2$ parameters.

For a $Z^\prime$ boson mass ranging in the interval 1 TeV $<m_{Z^\prime}<$ 2 TeV the $h_2(m_{Z^\prime})$ function is suppressed with respect to the $h_1(m_{Z^\prime})$ function about six orders of magnitude. Thus, the interference term, $h_2(m_{Z^\prime})(|Q^{e}_{L}\Omega_{R\mu\tau}|^2
+|Q^{e}_{R}\Omega_{L\mu\tau}|^2)$, can be neglected, which simplifies the calculation of $|\Omega_{L\mu\tau}|^2$ and $|\Omega_{R\mu\tau}|^2$ parameters. Therefore, the Eq.~(\ref{br3}) becomes
\begin{equation}\label{br3-1}
Br(\tau\to\mu\mu^+\mu^-)=\frac{g_2^2}{384 \pi^3}h_1(m_{Z^\prime})(|Q^{e}_{L}\Omega_{L\mu\tau}|^2+|Q^{e}_{R}\Omega_{R\mu\tau}|^2)\,\frac{m_\tau}{\Gamma_\tau}.
\end{equation}

\section{Numerical results and discussion}
Let us now introduce the numerical computations for different cases in which we can bound the $\mathrm{Re}(\Omega_{L\mu\tau}\Omega_{R\mu\tau}^\ast)$, $|\Omega_{L\mu\tau}|^2$, $|\Omega_{R\mu\tau}|^2$, and $|\Omega_{L\mu\tau}|^2+|\Omega_{R\mu\tau}|^2$ parameters, which represent the strength of the $Z^\prime\tau\mu$ coupling. We use the most restrictive bounds to predict the branching ratio of the $Z^\prime\to\tau\mu$ decay in the context of various extended models mentioned above.

\subsection{Bound according to the muon anomalous magnetic dipole moment result}
\begin{figure}[htb!]
\centering
\includegraphics[width=3.7in]{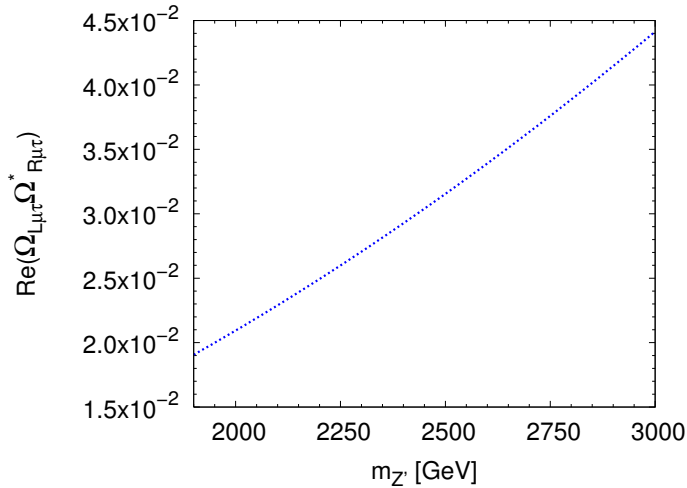}
\caption{\label{B-mdm} Behavior of $\mathrm{Re}(\Omega_{L\mu\tau}\Omega_{R\mu\tau}^\ast)$ as a function of the $Z^\prime$ boson mass.}
\end{figure}

From the experimental uncertainty of the anomalous magnetic dipole moment measurement and using Eq.~(\ref{amp91}) we bound the flavor-violating $\mathrm{Re}(\Omega_{L\mu\tau}\Omega_{R\mu\tau}^\ast)$ parameter as a function of the $Z^\prime$ boson mass. In Fig.~\ref{B-mdm} we can appreciate the behavior of the maximum of $\mathrm{Re}(\Omega_{L\mu\tau}\Omega_{R\mu\tau}^\ast)$ as a function of the $Z^\prime$ boson mass. Notice that the growth ranges from $10^{-3}$ to $10^{-2}$ for a $Z^\prime$ mass interval 1 TeV $< m_{Z^\prime} <$ 2 TeV.

As to the flavor-violating $\mathrm{Im}(\Omega_{L\mu \tau}\,\Omega_{R\mu \tau}^{\ast})$ parameter concerns, which is computed from the experimental limit on the muon electric dipole moment and using Eq.~(\ref{amp92}), gives a rather bad constraint since its least value is of the order of unity for a $Z^\prime$ mass interval 1 TeV $< m_{Z^\prime} <$ 2 TeV.
\subsection{Bounds according to the two and three body decays}

\subsubsection{Vector-like coupling}
\begin{figure}[htb!]
\centering
\includegraphics[width=3.7in]{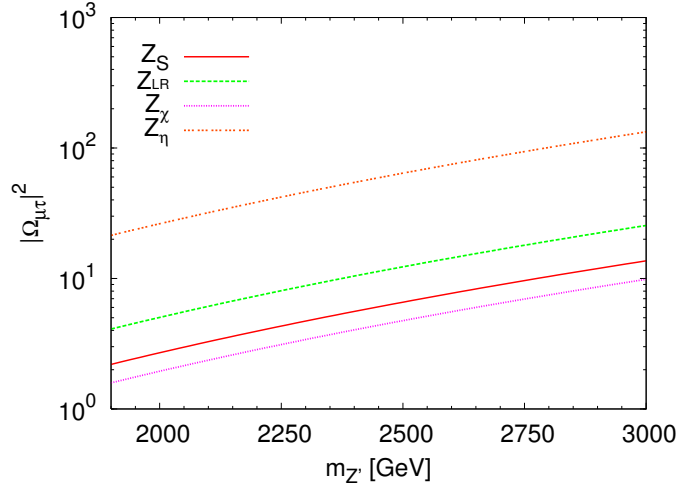}
\caption{\label{2-v} Behavior of $|\Omega_{\mu\tau}|^2$ as a function of the $Z^\prime$ boson mass. This graph corresponds to the bound obtained from the $\tau\to\mu\gamma$ decay.}
\end{figure}

As to the two-body $\tau\to\mu\gamma$ decay concerns, considering a vector-like coupling implies that $\Omega_{L\mu\tau}=\Omega_{R\mu\tau}$, for simplicity, in this case we define $\Omega_{L\mu\tau}\equiv \Omega_{\mu\tau}$. Solving for $|\Omega_{\mu\tau}|^2$ from Eq.~(\ref{br-2}) we can write the following inequality
\begin{equation}\label{bound-2-v}
|\Omega_{\mu\tau}|^2<\frac{4096 \pi^4}{\alpha}\frac{\Gamma_\tau}{m_\tau}\,\frac{\mathrm{Br_{Exp}}(\tau\to\mu\gamma)}{T_1},
\end{equation}
where \begin{equation}
T_1=g_2^2\,[|F_1(Q^e_L-Q^e_R)+F_2 Q^e_L|^2+|F_1(Q^e_R-Q^e_L)+F_2Q^e_R|^2].\nonumber
\end{equation}
The graph in Fig.~\ref{2-v} shows the behavior of the maximum of $|\Omega_{\mu\tau}|^2$ parameter provided by the inequality in Eq.~(\ref{bound-2-v}). We observe that the $Z_\chi$ case offers the most restrictive constraint, which ranges from $1$ to $10$ in the $Z^\prime$ boson mass interval 1.9 TeV $< m_{Z^\prime} <$ 3 TeV. In order to compare the arising bounds from the different models, it is sufficient to use only a value for a mass of the $Z^\prime$ boson since the different curves in Fig.~\ref{2-v} are all monotonically increasing. Let us consider the $T_1$ factor in Eq.~(\ref{bound-2-v}) for the different models with $m_{Z^\prime}=2$ TeV. For the $Z_\eta$ case $T_1=1.17453\times10^{-13}$, for the $Z_{LR}$ case $T_1=6.13335\times10^{-13}$, for the $Z_S$ case $T_1=1.14458\times10^{-12}$, and finally for the $Z_\chi$ case $T_1=1.58561\times10^{-12}$. From these numbers we may infer that the $Z_\chi$ case offers the most restrictive bound.

\begin{figure}[htb!]
\centering
\includegraphics[width=3.7in]{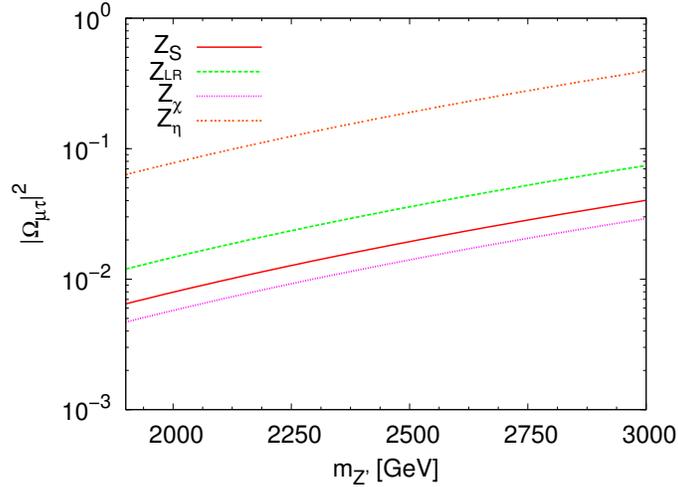}
\caption{\label{3-v} Behavior of $|\Omega_{\mu\tau}|^2$ as a function of the $Z^\prime$ boson mass. This graph corresponds to the bound obtained from the $\tau\to\mu\mu^+\mu^-$ decay.}
\end{figure}

In relation to the three-body $\tau\to\mu\mu^+\mu^-$ decay, taking a vector-like $Z^\prime$ boson and after solving for the $|\Omega_{\mu\tau}|^2$ parameter in Eq.~(\ref{br3-1}) we reach to the following inequality
\begin{equation}\label{bound-3-v}
|\Omega_{\mu\tau}|^2<384\,\pi^3
\frac{\Gamma_\tau}{m_\tau}\,\frac{\mathrm{Br_{Exp}}(\tau\to\mu\mu^+\mu^-)}{T_2\,h_1(m_{Z^\prime})},
\end{equation}
where
\begin{equation}
T_2=g_2^2({Q^e_L}^2+{Q^e_R}^2).\nonumber
\end{equation}
In Fig.~\ref{3-v} it can be appreciated the maximum of the $|\Omega_{\mu\tau}|^2$ parameter derived from Eq.~(\ref{bound-3-v}). We note that its behavior is monotone increasing for all models analyzed. Let us emphasize that the $Z_\chi$ case gives the most restrictive constraint, which is about three orders of magnitude less than the respective previous bound derived from the $\tau\to\mu\gamma$ decay. The strength of the $|\Omega_{\mu\tau}|^2$ parameter varies from $10^{-3}$ to $10^{-2}$ for a $Z^\prime$ boson mass within the range 1.9 TeV $< m_{Z^\prime} <$ 3 TeV. The $T_2$ factor in Eq.~(\ref{bound-3-v}) determines the difference between the bounds coming from the analyzed models. For the $Z_\eta$ case $T_2=6.62832\times10^{-3}$, for the $Z_{LR}$ case $T_2=3.50795\times10^{-2}$, for the $Z_{S}$ case $T_2=6.48551\times10^{-2}$, and for the $Z_\chi$ case $T_2=8.94823\times10^{-2}$. From this analysis we conclude that the $Z_\chi$ case provides the strongest bound.

\subsubsection{Maximal parity violation coupling}
\begin{figure}[htb!]
\centering
\includegraphics[width=3.5in]{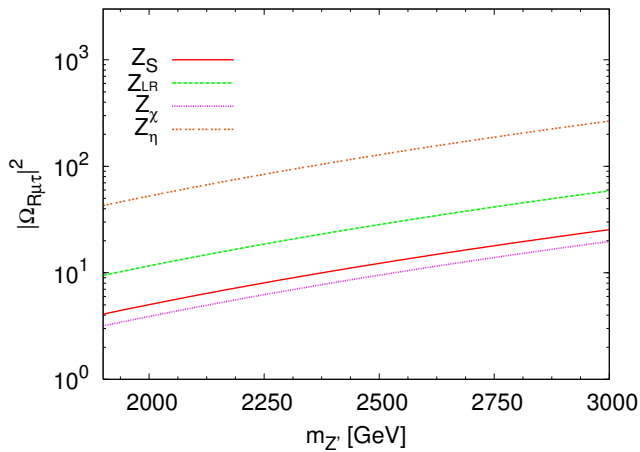}
\caption{\label{2-d} Behavior of $|\Omega_{R\mu\tau}|^2$ as a function of the $Z^\prime$ boson mass. This graph corresponds to the bound obtained from the $\tau\to\mu\gamma$ decay.}
\end{figure}

Returning to the case of the $\tau\to\mu\gamma$ decay, considering a vector-axial coupling, for simplicity, we take $\Omega_{L\mu\tau}=0$. Solving for $|\Omega_{R\mu\tau}|^2$ from Eq.~(\ref{br-2}) we can write the following inequality
\begin{equation}\label{bound-2-d}
|\Omega_{R\mu\tau}|^2<\frac{4096 \pi^4}{\alpha}\frac{\Gamma_\tau}{m_\tau}\,\frac{\mathrm{Br_{Exp}}(\tau\to\mu\gamma)}{T_3},
\end{equation}
where
\begin{equation}
T_3=g_2^2\,|F_1(Q^e_R-Q^e_L)+F_2Q^e_R|^2.\nonumber
\end{equation}
The graph in Fig.~\ref{2-d} shows the behavior of the maximum of $|\Omega_{R\mu\tau}|^2$ parameter provided by the inequality in Eq.~(\ref{bound-2-d}). We observe that the $Z_\chi$ case offers the most restrictive constraint, which ranges from $2$ to $20$ in the $Z^\prime$ boson mass interval 1.9 TeV $< m_{Z^\prime} <$ 3 TeV. To compare the arising bounds from the different models, it is sufficient to use only a value for a mass of the $Z^\prime$ boson since the different curves in Fig.~\ref{2-d} are all monotonically increasing. For $m_{Z^\prime}=2$ TeV, the $T_3$ factor in Eq.~(\ref{bound-2-d}) determines the discrepancies among the bounds of the different models studied. For the $Z_\eta$ case $T_3=5.87263\times10^{-14}$, for the $Z_{LR}$ case $T_3=2.65648\times10^{-13}$, for the $Z_S$ case $T_3=6.14351\times10^{-13}$, and for the $Z_\chi$ case $T_3=7.92805\times10^{-13}$. From last analysis it is evident that the $Z_\chi$ case gives the most restrictive bound.

\begin{figure}[htb!]
\centering
\includegraphics[width=3.5in]{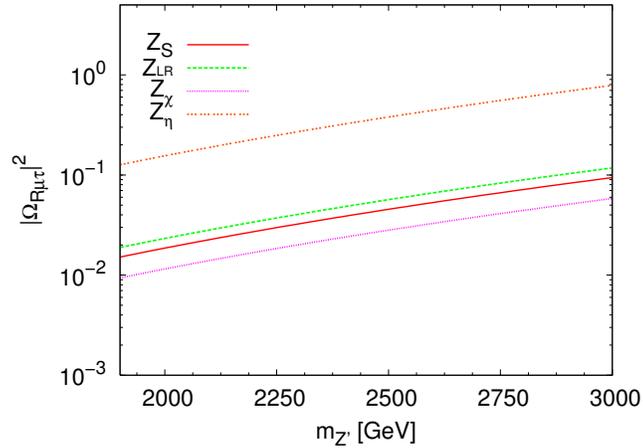}
\caption{\label{3-d} Behavior of $|\Omega_{R\mu\tau}|^2$ as a function of the $Z^\prime$ boson mass. This graph corresponds to the bound obtained from the $\tau\to\mu\mu^+\mu^-$ decay.}
\end{figure}

In relation to the three-body $\tau\to\mu\mu^+\mu^-$ decay, taking a vector-axial coupling and after solving for the $|\Omega_{R\mu\tau}|^2$ parameter in Eq.~(\ref{br3-1}) we reach to the following inequality
\begin{equation}\label{bound-3-d}
|\Omega_{R\mu\tau}|^2<384\,\pi^3\,\frac{\Gamma_\tau}{m_\tau}
\,\frac{\mathrm{Br_{Exp}}(\tau\to\mu\mu^+\mu^-)}{T_4\,h_1(m_{Z^\prime})},
\end{equation}
where
\begin{equation}
T_4=g_2^2{Q^e_R}^2.\nonumber
\end{equation}
In Fig.~\ref{3-d} we show the maximum of the $|\Omega_{R\mu\tau}|^2$ parameter derived from the inequality in Eq.~(\ref{bound-3-d}). We note that its behavior is monotone increasing for all models analyzed. Again the $Z_\chi$ case gives the most restrictive constraint, which is about two orders of magnitude less than the respective previous bound derived from the $\tau\to\mu\gamma$ decay. The strength of the $|\Omega_{R\mu\tau}|^2$ parameter varies from $10^{-3}$ to $10^{-2}$ for a $Z^\prime$ boson mass within the range 1.9 TeV $< m_{Z^\prime} <$ 3 TeV. The $T_4$ factor in Eq.~(\ref{bound-3-d}) determines the difference between the bounds coming from the analyzed models. For the $Z_\eta$ case $T_4=3.31416\times10^{-3}$, for the $Z_{LR}$ case $T_4=2.21696\times10^{-2}$, for the $Z_{S}$ case $T_4=2.76799\times10^{-2}$, and for the $Z_\chi$ case $T_4=4.47412\times10^{-2}$. The analysis evidences that the $Z_\chi$ case again gives the strongest bound.

\subsubsection{General coupling}
Here we analyze the most general bound, $|\Omega_{L\mu\tau}|^2+|\Omega_{R\mu\tau}|^2$, which can be constrained from the $\tau\to\mu\gamma$ and $\tau\to\mu\mu^+\mu^-$ decays. Notice that $|\Omega_{L\mu\tau}|^2+|\Omega_{R\mu\tau}|^2$ parameter can only be obtained in this particular combination for the $Z_\psi$ gauge boson derived from the $E_6$ model, since the $|Q^e_{L,R}|$ diagonal couplings satisfy the special relation $|Q^e_{L}|=|Q^e_{R}|$.

\begin{figure}[htb!]
\centering
\includegraphics[width=3.5in]{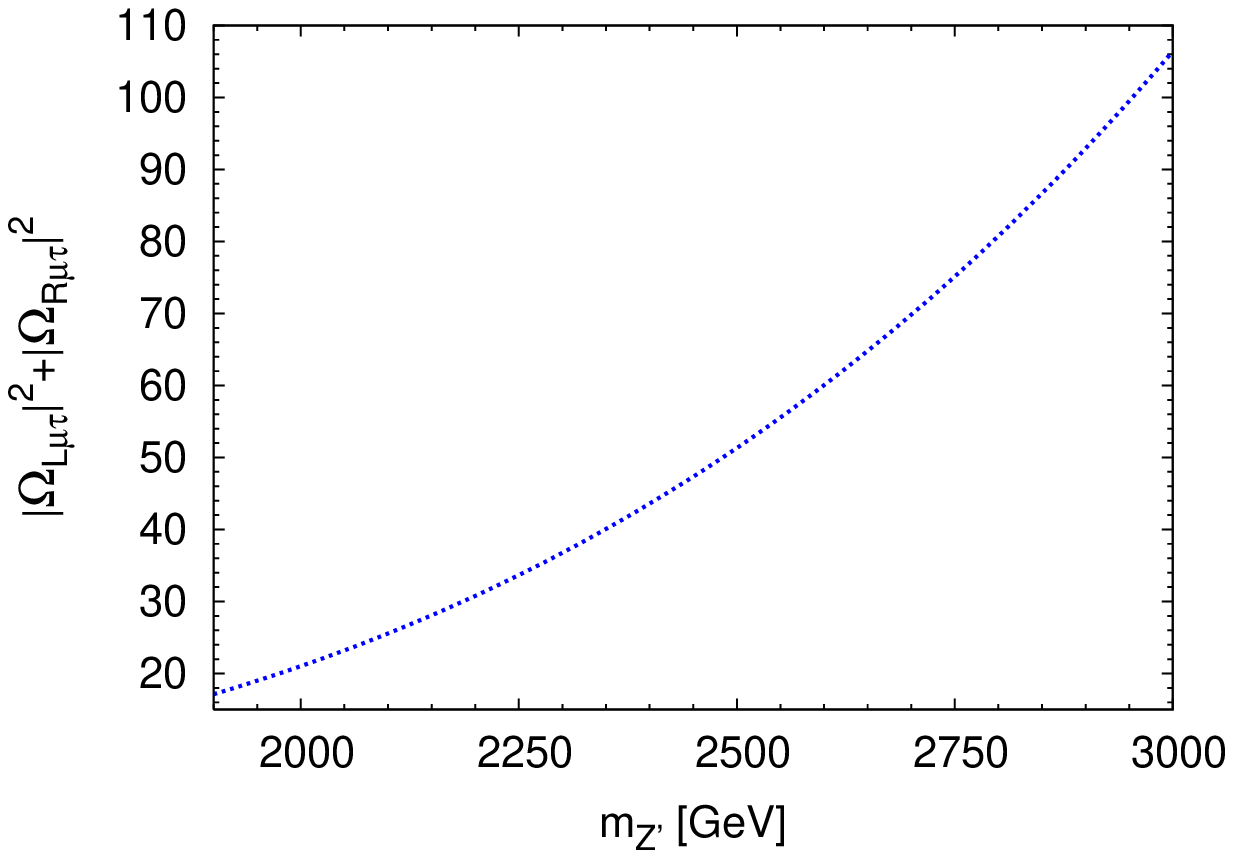}
\caption{\label{2-psi} Behavior of $|\Omega_{L\mu\tau}|^2+|\Omega_{R\mu\tau}|^2$ as a function of the $Z^\prime$ boson mass. This graph corresponds to the bound obtained from the $\tau\to\mu\gamma$ decay.}
\end{figure}

Regarding the two-body $\tau\to\mu\gamma$ decay, from Eq.~(\ref{br-2}) we can write the following inequality
\begin{equation}\label{bound-2-psi}
|\Omega_{L\mu\tau}|^2+|\Omega_{R\mu\tau}|^2<\frac{4096 \pi^4}{\alpha g_2^2}\frac{\Gamma_\tau}{m_\tau}\frac{\mathrm{Br_{Exp}}(\tau\to\mu\gamma)}{|2F_1+F_2|^2\,|Q^{e}_{L}|^2}.
\end{equation}
The graph in Fig.~\ref{2-psi} shows the behavior of the maximum of $|\Omega_{L\mu\tau}|^2+|\Omega_{R\mu\tau}|^2$ parameter provided by the inequality in Eq.~(\ref{bound-2-psi}). The intensity of the $|\Omega_{L\mu\tau}|^2+|\Omega_{R\mu\tau}|^2$ parameter ranges from $16$ to $106$ in the $Z^\prime$ boson mass interval 1.9 TeV $< m_{Z^\prime} <$ 3 TeV.

\begin{figure}[htb!]
\centering
\includegraphics[width=3.5in]{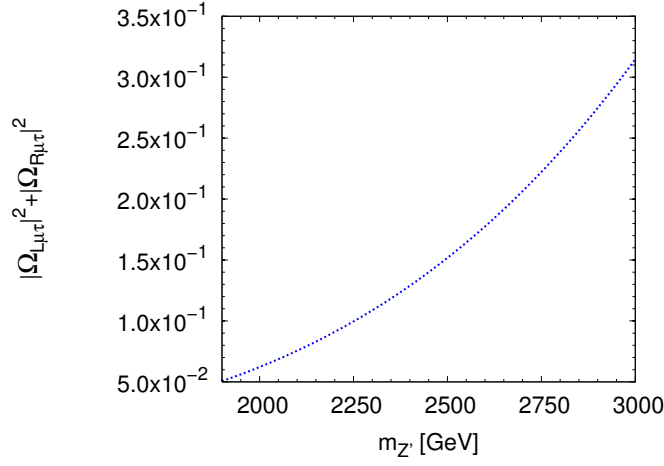}
\caption{\label{3-psi} Behavior of $|\Omega_{L\mu\tau}|^2+|\Omega_{R\mu\tau}|^2$ as a function of the $Z^\prime$ boson mass. This graph corresponds to the bound obtained from the $\tau\to\mu\mu^+\mu^-$ decay.}
\end{figure}

In relation to the three-body $\tau\to\mu\mu^+\mu^-$ decay, after solving for the $|\Omega_{L\mu\tau}|^2+|\Omega_{R\mu\tau}|^2$ parameter in Eq.~(\ref{br3-1}) we can write the following inequality
\begin{equation}\label{bound-3-psi}
|\Omega_{L\mu\tau}|^2+|\Omega_{R\mu\tau}|^2<\frac{384\pi^3}{g_2^2{Q^e_L}^2}
\frac{\Gamma_\tau}{m_\tau}\frac{\mathrm{Br_{Exp}}(\tau\to\mu\mu^+\mu^-)}{h_1(m_{Z^\prime})}.
\end{equation}
In Fig.~\ref{3-psi} it can be appreciated the maximum of the $|\Omega_{L\mu\tau}|^2+|\Omega_{R\mu\tau}|^2$ parameter as a function of the $Z^\prime$ boson mass provided by the inequality in Eq.~(\ref{bound-3-psi}). We note that its behavior is monotone increasing for the $Z_\psi$ case. The strength of the $|\Omega_{L\mu\tau}|^2+|\Omega_{R\mu\tau}|^2$ parameter varies from $10^{-2}$ to $10^{-1}$ for a $Z^\prime$ boson mass within the range 1.9 TeV $< m_{Z^\prime} <$ 3 TeV. Let us emphasize that this constraint is about three orders of magnitude less than the respective previous bound derived from the $\tau\to\mu\gamma$ decay.

\subsection{The branching ratio of the $Z^\prime\to\tau\mu$ process}

In order to make predictions, we resort to the bounds obtained from the previous analysis. As discussed above, we are interested in studying the $Z'\to\tau\mu$ decay, whose branching ratio can be written as
\begin{align}
\mathrm{Br}(Z^\prime\to \tau\mu)=\frac{1}{24 \pi}\frac{m_{Z^\prime}}{\Gamma_{Z^\prime}}&\Bigg[\Bigg(1-\frac{(m_\tau+m_\mu)^2}{m^2_{Z^\prime}}\Bigg)
\Bigg(1-\frac{(m_\tau-m_\mu)^2}{m^2_{Z^\prime}}\Bigg)\Bigg]^{1/2}\Bigg[\Bigg(2-\frac{m_\tau^2+m_\mu^2}{m_{Z^\prime}^2}
-\frac{(m_\tau^2-m_\mu^2)^2}{m_{Z^\prime}^4}\Bigg)\nonumber\\
&\times (|\Omega_{L\mu\tau}|^2+|\Omega_{R\mu\tau}|^2)+12\frac{m_\tau m_\mu}{m_{Z^\prime}^2}\mathrm{Re}(\Omega_{L\mu\tau}\Omega_{R\mu\tau}^\ast)\Bigg].
\end{align}
where $\Gamma_{Z'}$ is the $Z'$ total decay width. On account $\Gamma_{Z'}$ we include the total possible flavor-conserving and flavor-violating decay modes~\cite{arhrib, arandaetal}, namely $\nu_e\bar\nu_e$, $\nu_\mu\bar\nu_\mu$, $\nu_\tau\bar\nu_\tau$, $e\bar e$, $\mu\bar\mu$, $\tau\bar\tau$, $u\bar u$, $c\bar c$, $t\bar t$, $d\bar d$, $s\bar s$, $b\bar b$, $\bar uc+u\bar c$, $\bar tc+t\bar c$, and $\bar\tau\mu+\tau\bar\mu$. To compute the branching ratios, we will use the most restrictive bounds for the $|\Omega_{L\mu\tau}|^2$, $|\Omega_{R\mu\tau}|^2$, and $|\Omega_{L\mu\tau}|^2+|\Omega_{R\mu\tau}|^2$ parameters.

\begin{figure}[htb!]
\centering
\includegraphics[width=3.4in]{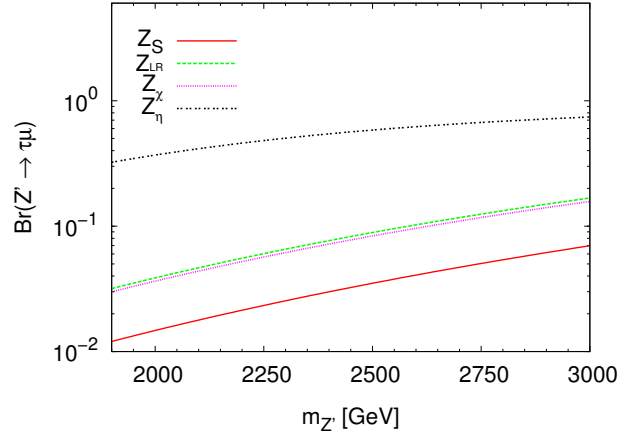}
\caption{\label{ratios-p-v} The branching ratio for the $Z^{\prime}\to\tau\mu$ as function of the $Z^{\prime}$ boson mass. This graph corresponds to a vector-like $Z^\prime\mu\tau$ coupling.}
\end{figure}

In Fig.~\ref{ratios-p-v} we show the branching ratio for the $Z'\to\tau\mu$ decay in the context of the sequential $Z$ model, the left-right symmetric model, the $SO(10)\rightarrow SU(5)\times U(1)$ model, and the superstring-inspired models in which $E_6$ breaks to a rank-5 group~\cite{langacker-rmp, langacker2}. From this figure, it is more feasible to observe lepton flavor violation for a $Z_\eta$ gauge boson, since the related branching ratio can be as higher as $7.4\times 10^{-1}$, while the more restrictive branching ratio corresponds to the sequential model, in which its strength can be as higher as $6.97\times 10^{-2}$.

\begin{figure}[htb!]
\centering
\includegraphics[width=3.4in]{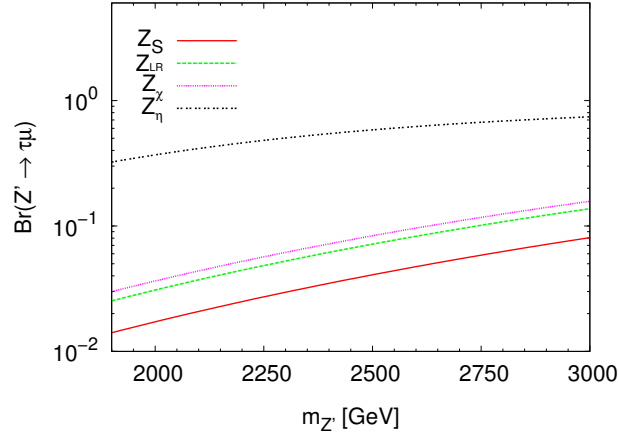}
\caption{\label{ratios-p-d} The branching ratio for the $Z^{\prime}\to\tau\mu$ as function of the $Z^{\prime}$ boson mass. This graph corresponds to a vector-axial $Z^\prime\mu\tau$ coupling.}
\end{figure}

The graph in the Fig.~\ref{ratios-p-d} shows the branching ratio for the $Z'\to\tau\mu$ decay in the context of the sequential $Z$ model, the left-right symmetric model, the $SO(10)\rightarrow SU(5)\times U(1)$ model, and the superstring-inspired models in which $E_6$ breaks to a rank-5 group~\cite{langacker-rmp, langacker2}. From this graph, again is more feasible to observe lepton flavor violation for a $Z_\eta$ gauge boson, since the related branching ratio can be as higher as $7.4\times 10^{-1}$. However, we can observe that the branching ratio for the $Z_\chi$ gauge boson has a similar strength, which make difficult to say, from a experimental point of view, to what $Z^\prime$ model corresponds. In contrast, the more restrictive bound for $\mathrm{Br}(Z^\prime\to\tau\mu)$ corresponds to the sequential model, in which the respective numerical value can be as higher as $8\times 10^{-2}$.

\begin{figure}[htb!]
\centering
\includegraphics[width=3.4in]{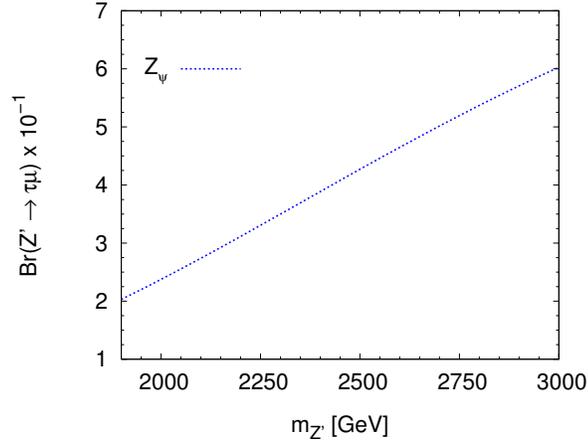}
\caption{\label{ratios-p-psi} The branching ratio for the $Z^{\prime}\to\tau\mu$ as function of the $Z^{\prime}$ boson mass. This graph corresponds to a $Z_\psi$ gauge boson.}
\end{figure}

In Fig.~\ref{ratios-p-psi} we show the branching ratio for the $Z'\to\tau\mu$ decay in the context of the $E_6\rightarrow SO(10)\times U(1)$ model~\cite{langacker-rmp, langacker2}. From this figure, we observe that the branching ratio can be as higher as $6\times 10^{-1}$. This is the more intense branching ratio for all models studied, which implies that lepton flavor violation for a $Z^\prime$ gauge boson in the context of the $E_6\rightarrow SO(10)\times U(1)$ model has high probability of being observed.

Let us estimate the discovery potential of $\tau\mu$ lepton flavor violation mediated by $Z^\prime$ boson at present and future in LHC. Here, we use a $Z'$ boson mass of 2 TeV according to the current lower limit imposed by CMS collaboration~\cite{cms1a}. The expected number of events, $N$, is given by $N=\mathcal{L}\,\sigma(pp\to Z')\,\mathrm{Br}(Z'\to\tau\mu)$, where $\mathcal{L}$ is the integrated luminosity of LHC, $\sigma(pp\to Z')$ is the $Z'$ production cross section at LHC which was calculated in a recent work~\cite{cp-yuan}, and the corresponding branching ratio is $1.47\times 10^{-2}$. Currently, the most conservative integrated luminosity corresponds to ATLAS detector~\cite{current-luminosity}, which is $\mathcal{L}=6.57\,\, \mathrm{fb}^{-1}$ at 4 TeV LHC. Therefore, we may assume $\mathcal{L}\approx10\,\,\mathrm{fb}^{-1}$ at 7 TeV energy collision, where $\sigma(pp\to Z')\approx 40$ fb~\cite{cp-yuan}, which implies that $N\approx 6$ events. In the future, for a long run it would be feasible to reach an integrated luminosity $\mathcal{L}\approx100\,\,\mathrm{fb}^{-1}$ at 14 TeV LHC. At this energy collision $\sigma(pp\to Z')\approx 700$ fb~\cite{cp-yuan} and we obtain $N\approx 1029$ events.

\section{Final remarks}
The current experimental data  by the CMS collaboration evidence the possibility of the existence of the $Z^{\prime}$ gauge boson which is predicted in several theories. Whether this boson exits, it could give a solution of the disagreement between the experimental results and the corresponding predicted by the SM  on lepton flavor violating transitions. In this work we employed the most general renormalizable Lagrangian that includes lepton flavor violation mediated by a new neutral massive gauge boson to bound the $Z^{\prime}\tau\mu$ coupling in the spirit of a model-independent approach.  We used the  experimental uncertainty of the muon magnetic dipole moment, which is measured with great precision, to obtain the maximum of $\mathrm{Re}(\Omega_{L\mu\tau}\Omega_{R\mu\tau}^\ast)$  as a function of the $Z^\prime$ boson mass, which is of the order of $10^{-2}$ for $1.9\;\mathrm{TeV}<m_{Z^\prime}<3\;\mathrm{TeV}$.
Additionally, we also perform a calculation of the $\tau\to\mu\gamma$ and $\tau\to\mu\mu^+\mu^-$ transitions and use their respective experimental restrictions in order to bound the $Z^{\prime}\tau\mu$ coupling. Specifically, we compute the maximum of the $|\Omega_{L\mu\tau}|^2$, $|\Omega_{R\mu\tau}|^2$, and $|\Omega_{L\mu\tau}|^2+|\Omega_{R\mu\tau}|^2$ parameters as a function of the $Z^\prime$ boson mass. We study three different cases: vector-like coupling, vector-axial coupling, and general coupling. We found that the most restrictive constraints corresponds to that coming from the $\tau\to\mu\mu^+\mu^-$ decay. From this decay, the most restrictive bound corresponds to the vector-like coupling, which is determined in the context of the sequential $Z$ model, where $|\Omega_{\mu\tau}|^2$ ranges from $10^{-3}$ to $10^{-2}$ for $1.9\;\mathrm{TeV}<m_{Z^\prime}<3\;\mathrm{TeV}$.

To calculate the $Z^{\prime}\to\mu\tau$ branching ratio we employ the most restrictive bounds for the $|\Omega_{L\mu\tau}|^2$, $|\Omega_{R\mu\tau}|^2$, and $|\Omega_{L\mu\tau}|^2+|\Omega_{R\mu\tau}|^2$ parameters. Among the analyzed models ($Z_S$, $Z_{LR}$, $Z_\chi$, $Z_\psi$, and $Z_{\eta}$) we found that the most restrictive branching ratio corresponds to the Sequential model in which $\mathrm{Br}(Z^{\prime}\to\mu\tau)$ is of the order of $10^{-2}$ for a $Z^\prime$ boson mass in the interval $1.9\;\mathrm{TeV}<m_{Z^\prime}<3\;\mathrm{TeV}$.

Surprisingly the most restrictive values for $\mathrm{Br}(Z^{\prime}\to\tau\mu)$ are of the order of $10^{-2}$, in a broad range of the $Z^{\prime}$ gauge boson mass studied. This relaxed bound enables the possibility of LFV being observed by $Z^\prime$ mediated processes. Our results show that current experimental constraints on lepton flavor violation restrict weakly this phenomenon mediated by an extra $Z$ gauge boson. In fact, at present for a $Z'$ boson mass of 2 TeV we estimated around 6 events associated with the $Z'\to\tau\mu$ process for 7 TeV LHC. In the future, for the same $Z'$ boson mass we obtained around 1029 events for 14 TeV LHC.

\section*{Acknowledgments}
This work has been partially supported by CONACYT and CIC-UMSNH.


\begin{thebibliography}{99}

\bibitem{Franceschini} Roberto Franceschini, Pier Paolo Giardino, Gian F. Giudice, Paolo Lodone, and Alessandro Strumia, JHEP \textbf{05}, 092 (2011).

\bibitem{cms2} S. V. Shmatov [CMS Collaboration], Phys. Atom. Nucl. \textbf{74}, 490 (2011), Yad. Fiz. \textbf{74}, 511 (2011).

\bibitem{ellis1}J.~Ellis, G.~Giudice, M.~Mangano, I.~Tkachev, and U.~Wiedemann, J. Phys. G: Nucl. Part. Phys. 35, 115004 (2008).

\bibitem{atlas1} G.~Aad {\it et al.} [ATLAS Collaboration], Phys.\ Lett.\ B\ {\bf 705}, 174  (2011).

\bibitem{cms1} S. Chatrchyan \emph{et al.} [CMS Collaboration], Phys. Rev. Lett. \textbf{106}, 231801 (2011).

\bibitem{cms1a} S.~Chatrchyan {\it et al.} [CMS Collaboration], JHEP\ {\bf 05}, 093  (2011); The CMS Collaboration, report CMS PAS EXO-11-019.

\bibitem{cms3} S.~Chatrchyan {\it et al.} [CMS Collaboration], Phys.\ Lett.\ B\ {\bf 704}, 123 (2011).

\bibitem{cms4} The CMS Collaboration, \textit{Search for BSM $t\bar{t}$ Production in the Boosted All-Hadronic Final State}, report CMS PAS EXO-11-006.

\bibitem{atlas2} The ATLAS Collaboration, report ATLAS-CONF-2012-007.

\bibitem{pleitez} F. Pisano and V. Pleitez, Phys. Rev. D \textbf{46}, 410 (1992); P. H. Frampton, Phys. Rev. Lett. \textbf{69}, 2889 (1992).

\bibitem{langacker1} M. Cveti\v{c}, P. Langacker, and B. Kayser, Phys. Rev. Lett. \textbf{68}, 2871 (1992); M. Cveti\v{c} and P. Langacker, Phys. Rev. D \textbf{54}, 3570 (1996); M. Cveti\v{c}  \emph{et al.}, Phys. Rev. D \textbf{56}, 2861 (1997); \textit{ibid}. \textbf{58}, 119905(E) (1998); M. Masip and A. Pomarol, Phys. Rev. D \textbf{60}, 096005 (1999); N. Arkani-Hamed, A. G. Cohen, and H. Georgi, Phys. Lett. B \textbf{513}, 232 (2001); N. Arkani-Hamed, A. G. Cohen, E. Katz, and A. E. Nelson, JHEP \textbf{07}, 034 (2002); T. Han, H. E. Logan, B. McElrath, and L.-T. Wang, Phys. Rev. D \textbf{67}, 095004 (2003); C. T. Hill and E. H. Simmons, Phys. Rept. \textbf{381}, 235 (2003); \emph{ibid}. \textbf{390}, 553 (2004); J. Kang and P. Langacker, Phys. Rev. D \textbf{71}, 035014 (2005); B. Fuks \emph{et al.}, Nucl. Phys. B \textbf{797}, 322 (2008); J. Erler \emph{et al.}, JHEP \textbf{08}, 017 (2009); M. Goodsell \emph{et al.}, JHEP \textbf{11}, 027 (2009); P. Langacker, AIP Conf. Proc. \textbf{1200}, 55 (2010).

\bibitem{perez-soriano} M. A. Perez and M. A. Soriano, Phys. Rev. D \textbf{46}, 284 (1992).

\bibitem{langacker-rmp} P. Langacker, Rev. Mod. Phys. \textbf{81}, 1199 (2009).

\bibitem{leike}  A. Leike, Phys. Rept. \textbf{317}, 143 (1999).

\bibitem{robinett} R. W. Robinett and Jonathan L. Rosner, Phys. Rev. D \textbf{26}, 2396 (1982).

\bibitem{langacker2} P. Langacker and M. Luo, Phys. Rev. D \textbf{45}, 278 (1992).

\bibitem{arhrib} A. Arhrib, \emph{et al.}, Phys. Rev. D \textbf{73}, 075015 (2006).

\bibitem{arandaetal} J. I. Aranda, F. Ram\'irez-Zavaleta, J. J. Toscano, and E. S. Tututi,  J. Phys. G: Nucl. Part. Phys. \textbf{38}, 045006 (2011).


\bibitem{neutrinos} R. Becker-Szendy \emph{et al}., Nucl. Phys. Proc. Suppl. \textbf{38}, 331 (1995);

  Y. Fukuda \emph{et al}., Phys. Lett. B \textbf{335}, 237 (1994); Phys. Rev. Lett. \textbf{81}, 1562 (1998); H. Sobel, Nucl. Phys. Proc. Suppl. \textbf{91}, 127 (2001); M. Ambrossio \emph{et al}., Phys. Lett. B \textbf{566}, 35 (2003); M. Apollonio \emph{et al}., Eur. Phys. J. \textbf{C 27}, 331 (2003); M. B. Smy \emph{et al}., Phys. Rev. D \textbf{69}, 011104(R) (2004); S. N. Ahmed \emph{et al}., Phys. Rev. Lett. \textbf{92}, 181301 (2004); Y. Ashie \emph{et al}., Phys. Rev. Lett. \textbf{93}, 101801 (2004); E. Aliu \emph{et al}., Phys. Rev. Lett. \textbf{94}, 081802 (2005); Y. Ashie \emph{et al}., Phys. Rev. D \textbf{71}, 112005 (2005); W. W. M. Allison \emph{et al}., Phys. Rev. D \textbf{72}, 052005 (2005); P. Adamson \emph{et al}., Phys. Rev. D \textbf{73}, 072002 (2006).

\bibitem{cheng-li} T-P Cheng, L-F Li, \textit{Gauge Theory of Elementary Particle Physics},  Clarendon Press, Oxford (1984).

\bibitem{pdg} K. Nakamura \textit{et al}., J. Phys. G: Nucl. Part. Phys. \textbf{37}, 075021 (2010).

\bibitem{lfv} G. D'Ambrosio, G. F. Giudice, G. Isidori, and A. Strumia, Nucl. Phys. B \textbf{645}, 155 (2002); V. Cirigliano, B. Grinstein, G. Isidori, and M. B. Wise, Nucl. Phys. B \textbf{728}, 121 (2005); S. Davidson and F. Palorini, Phys. Lett. B \textbf{642}, 72 (2006); A. Mondragon, M. Mondragon, and E. Peinado, Phys. Rev. D \textbf{76}, 076003 (2007); R. Alonso, G. Isidori, L. Merlo, L. A. Mu\~noz, and E. Nardi, JHEP \textbf{06}, 037 (2011); L.~Merlo, S.~Rigolin, and B.~Zaldivar, JHEP\ {\bf 11}, 047 (2011);  $\mathrm{\dot{I}}$.~$\mathrm{\c{S}}$ahin and M.~K$\mathrm{\ddot{o}}$ksal, arXiv:1108.5363.

\bibitem{durkin} L. S. Durkin and P. Langacker, Phys. Lett. B \textbf{166}, 436 (1986);
 M. Cvetic and P. Langacker, Proceedings of Ottawa 1992:  Beyond the standard model 3, 454-458, (1992); Cheng-Wei Chiang, Yi-Fan Lin, and Jusak Tandean, JHEP 11, 083 (2011).
\bibitem{langacker3} P. Langacker and M. Pl$\mathrm{\ddot{u}}$macher, Phys. Rev. D \textbf{62}, 013006 (2000); X.-G. He and G. Valencia, Phys. Rev. D \textbf{74}, 013011 (2006); C.-W. Chiang, N. G. Deshpande, and J. Jiang, JHEP \textbf{08}, 075 (2006).
\bibitem{Salam-Mohapatra} J. C. Pati and A. Salam, Phys. Rev. D \textbf{10}, 275 (1974); R. N. Mohapatra and J. C. Pati, Phys. Rev. D \textbf{11}, 566 (1975).

\bibitem{Pleitez} F. Pisano and V. Pleitez, Phys. Rev. D \textbf{46}, 410 (1992); P. H. Frampton, Phys. Rev. Lett. \textbf{69}, 2889 (1992).

\bibitem{robinett2} Richard W. Robinett and Jonathan L. Rosner, Phys. Rev. D \textbf{25}, 3036 (1982); R. W. Robinett, Phys. Rev. D \textbf{26}, 2388 (1982).

\bibitem{babar} B.~Aubert {\it et al.}  [BABAR Collaboration], Phys.\ Rev.\ Lett.\  \textbf{104}, 021802 (2010).

\bibitem{belle} K.~Hayasaka {\it et al.} [BELLE Collaboration], Phys.\ Lett.\ B \textbf{687}, 139 (2010).

\bibitem{cp-yuan} Qing-Hong Cao, Zhao Li, Jiang-Hao Yu, and C. P. Yuan, arXiv:1205.3769 [hep-ph].

\bibitem{current-luminosity} The LHC Performance and Statistics, http://lhc-statistics.web.cern.ch/LHC-Statistics/.

\end{thebibliography}
\end{document}